\documentclass[preprint,aps,prb]{revtex4}
\usepackage{graphicx}
\usepackage{amssymb}

\newcommand{\eq}[1]{Eq.(\ref{#1})}

\begin{document}

\title{Liquid-vapour transition of the long range Yukawa fluid}

\author{Jean-Michel Caillol $^1$, Federica Lo Verso  $^2$,
  Elisabeth Sch\"oll-Paschinger  $^3$, Jean-Jacques Weis $^1$}
 
\affiliation{ $^1$Laboratoire de Physique Th\'eorique,
UMR 8627, B\^at. 210, Universit\'e de Paris-Sud,
91405 Orsay Cedex, France\\
  $^2$Institut f\"ur Theoretische Physik II,
Heinrich-Heine-Universit\"at D\"usseldorf,
Universit\"atsstrasse 1, D-40225 D\"usseldorf, Germany \\
  $^3$CMS and Fakult\"at f\"ur Physik, Universit\"at Wien, 
Boltzmanngasse 5, A-1090 Wien, Austria} 
\date{\today}

\begin{abstract}
Two liquid state theories, the  self-consistent Ornstein-Zernike
equation (SCOZA) and the hierarchical reference theory (HRT) are shown,
by comparison with Monte Carlo simulations, to perform extremely well in
predicting the liquid-vapour coexistence  of the hard core Yukawa (HCY) fluid when
the interaction is long range. The long range of the potential is
treated in the simulations using both an Ewald sum and hyperspherical
boundary conditions. In addition, we present an analytical optimised
mean field theory which is exact in the limit of an infinitely long
range interaction. The work extends a previous one by Caccamo {\em et
  al} [{\itshape Phys. Rev. E,} {\bfseries 60}, 5533 (1999)] for short range interactions.

\textbf{keywords} : Yukawa potential, Critical phenomena, Monte Carlo
simulations, SCOZA, HRT.\\

\end{abstract}
\maketitle
\section{\label{intro}Introduction}

Two liquid state theories are presently available which provide an accurate
description of the liquid-vapour transition of simple systems including the
critical region. A first one is the so-called self-consistent Ornstein-Zernike
approximation (SCOZA) devised initially by H{\o}ye and Stell
\cite{scoza:12,scoza:11} and later adapted for numerical calculations
\cite{scoza:1,scoza:8,scoza:9}. It is based on the assumption that the direct
correlation function behaves, to a good approximation, at long range
(attractive region of the potential) as the potential with a density and
temperature dependent prefactor which is determined by imposing consistency
between the energy and compressibility routes to the thermodynamics. The
second approach put forward by Parola and Reatto \cite{Parola:85},
incorporates renormalisation group ideas by gradually taking into account
fluctuations of longer and longer wavelengths. It is based on the hierarchical
reference theory (HRT) of fluids truncated at lowest order by means of an ORPA-like
(optimised random phase approximation) ansatz (see Ref.\ \cite{hrtrev} for
review).  
Both predict
non-classical critical exponents \cite{Parola:85,scoza:25}.  
Thermodynamic
properties and phase diagrams based on the SCOZA and HRT approaches have been
reported for a variety of potentials (limiting ourselves to continuum systems)
including the hard core Yukawa (HCY) potential
\cite{scoza:6,Caccam:99,Pini:00,Pini:06}, the square well potential
\cite{Reiner:02,scoza:26}, the Lennard-Jones (LJ) potential
\cite{mero,tau,Hoye:06,scoza06}, the Gaussian potential \cite{Mladek:06},
ultrasoft potentials \cite{soft-F},
Asakura-Oosawa pair potentials \cite{HRT-F,AO-FG}, Girifalco potentials~\cite{tau,SPKEPL,PRE03}, binary hard core
Yukawa mixtures \cite{Pini:03,SPKJCP,plwk:bin:symm,jcp05}, 
etc..
Most of the comparisons with simulation data are so far restricted to short
or medium range potentials. The aim of this article is to extend such
comparisons 
to long range interactions. A convenient choice to do this is the HCY
potential as the range can be varied from short range appropriate for modelling
colloidal suspensions or protein solutions to medium range, where it
mimics the familiar LJ potential \cite{Caccam:99}, to long
range. For long range interactions we have considered a
third approach, an optimised mean field theory (OMF) \cite{Cai-Mol},
exact in the Kac limit \cite{Kac:63}, which can be worked out
analytically for the HCY potential (see section 6).  
A further advantage of the HCY potential is that the repulsive part of the potential which in
SCOZA, HRT and OMF is used as a reference potential  is a hard sphere
potential whose properties are accurately known
\cite{Carnah:69,verlet,Hender:75}. Last, an approximate 
analytical solution (mean spherical approximation) is available for the
HCY potential \cite{Waisma:73,hoyeblum} which  drastically reduces the
computational efforts of the solution of the SCOZA equations.

For short inverse screening lengths $\alpha$ of the Yukawa potential,
$1.8 \leq \alpha\sigma=\alpha^*  < 7$ ($\sigma$ hard sphere diameter), a comparison
between SCOZA, HRT and simulation results for the liquid-vapour
coexistence curve is available in \cite{Caccam:99}.
In this work we extend the domain of  $\alpha$ 
 from $\alpha^*=1.8$ (where the HCY potential approximates
the LJ potential) to  $\alpha^*=0$, the infinite long range potential.
With respect to simulations, in this domain the range of the Yukawa
potential generally exceeds the dimensions of the simulation box (for
typical system sizes of a thousand particles) and some care is necessary
to properly treat the long range of the potential. This has been done
both  by using periodic boundary conditions (b.c.)
and   performing an Ewald (EW) sum \cite{Cai-Salin1}   and by
using hyperspherical b.c. \cite{Cai-Gilles,Cai-Gilles2}.    
 
The remainder of the paper is organised as follows: 
In section \ref{Mod} we define the HCY interaction. Section
\ref{Sim} provides a description of the two Monte Carlo (MC) methods used to treat the
long range of the Yukawa interaction, Ewald sum and hyperspherical
method.
After summarising the three considered theories, SCOZA, HRT and OMF in
section \ref{Th} we compare, in section \ref{results}, the numerical
results obtained with the different theories for the liquid-vapour
coexistence curve and the thermodynamic properties along the coexistence
boundary with the simulation results. The conclusions are presented in
section \ref{conclusion}. An appendix describes some properties of Yukawa charge
distributions relevant for elaborating the OMF theory.

\section{\label{Mod} Model}
In our system particles interact by the hard-core Yukawa potential

\begin{equation}
v(r) =  \left\{ \begin{array} {l@{~~~~~~~~}l}
                             \infty  & r \le \sigma \\
             \displaystyle  -q^2 \alpha^{*2} \frac{e^{-\alpha^* r/\sigma}}{r} 
                                     & r > \sigma \\
                             \end{array}
                             \right. .
\label{potential}
\end{equation}
In the limit $\alpha \to 0$ the potential gets infinitely long range and
infinitely weak (Kac limit).
We define an inverse 
reduced temperature  $\beta^*= 1/T^* = q^2/k_{\rm B}T\sigma$, where 
 $k_{\rm B}$ is the Boltzmann
constant, $T$ the temperature and $q$ has dimension of an electric charge. The
reduced density is $\rho^*= \rho \sigma^3$. 

\section{\label{Sim}Simulations.}

Numerical simulations of fluids or plasmas involving Yukawa interactions
$v(r) \propto \exp(-\alpha r)/r$ can be performed either in a cube of
side $L$ with periodic b.c. ($\mathcal{C}_{3}$) or on the
surface of a four dimensional (4D) sphere of centre $O$ and radius $R$
(equation $x^2 + y^2 + z^2 + u^2 =R^2$), the hypersphere  
$\mathcal{S}_{3}$ for short. In both geometries the Helmholtz equation
$(\Delta -\alpha^2)v(\bf{x})=-4 \pi \delta(\bf{x})$ can be solved
analytically.

\subsection{\label{Ew} Ewald sums.}
In the case of $\mathcal{C}_{3}$ the solution of the Helmholtz equation
can be reexpressed in terms of Ewald sums with good convergence
properties\cite{Cai-Salin1}. The use of an  Ewald potential
is of course only necessary when the range of the Yukawa potential
exceeds the linear dimensions of the simulation box.
 Molecular
dynamics simulations of simple models of Yukawa plasmas were performed
recently with this method\cite{Cai-Salin2,Cai-Salin3}.
For the potential model Eq. ~(\ref{potential}) 
the Ewald sum for the energy is given by
\cite{Cai-Salin1}
\begin{eqnarray}
\label{I2}
U =&& -\frac{1}{2} q^2 \alpha^{*2}  \sum_{i=1}^N \sum_{j=1}^N \sum_{\bf n} {}^{'}
  \left[ {\rm erfc} \Big(\eta |{\bf r}_{ij}+ L{\bf n}|+ \displaystyle
  \frac{\alpha}{2 \eta} \Big) e^{\alpha  |{\bf r}_{ij}+ L{\bf n}|} \right.
  \nonumber \\  
&& \left. + {\rm erfc} \Big(\eta |{\bf r}_{ij}+ L{\bf n}|-  \displaystyle
  \frac{\alpha}{2 \eta} \Big) e^{-\alpha  |{\bf r}_{ij}+ L{\bf n}|}
  \right] \Big /2  |{\bf r}_{ij}+ L{\bf n}| \nonumber \\
&& - \frac{2\pi}{V} q^2 \alpha^{*2} \sum_{{\bf k} =0}^\infty \displaystyle 
 \frac{e^{- ({\bf k}^2 +\alpha^2) / 4 \eta^2}}{{\bf k}^2 +\alpha^2}
  F({\bf k})F^*({\bf k}) +  U_{self} 
\end{eqnarray}
where
\begin{eqnarray}
\label{I3}
U_{self}/N =&& -\frac{1}{2}  q^2 \alpha^{*2} \sum_{\bf n \neq 0} 
  \left[ {\rm erfc} \Big(\eta L |{\bf n}|+ \displaystyle
  \frac{\alpha}{2 \eta} \Big) e^{\alpha L |{\bf n}|} \right.
  \nonumber \\  
&& \left. + {\rm erfc} \Big(\eta L |{\bf n}|-  \displaystyle
  \frac{\alpha}{2 \eta} \Big) e^{-\alpha L |{\bf n}|}
  \right] \Big /2 L |{\bf n}| \nonumber \\
&& - \frac{2\pi}{V} q^2 \alpha^{*2} \sum_{{\bf k} =0}^\infty \displaystyle 
 \frac{e^{- ({\bf k}^2 +\alpha^2) / 4 \eta^2}}{{\bf k}^2 +\alpha^2}
  \nonumber \\
&& + q^2 \alpha^{*2} \frac{\eta}{\sqrt \pi}  {e^{- \alpha^2 / 4 \eta^2}}
- q^2 \alpha^{*2} \; \frac{\alpha}{2} {\rm erfc} \Big( \displaystyle \frac{\alpha}{2
  \eta} \Big).  
\end{eqnarray}

and
\begin{eqnarray}
\label{I6}
F({\bf k}) &=& \sum_{i=1}^N  \displaystyle \exp[i{\bf k} \cdot {\bf r}_i],
\end{eqnarray}
In Eq.\ (\ref{I2}) ${\bf r}_{ij}={\bf r}_j-{\bf r}_i$,  
$L$ is the box length, $V=L^3$ the volume and $\rm{erfc}$ denotes the 
complementary error function. The prime in the sum over 
${\bf n} = (n_x,n_y,n_z)$, with $n_x$,$n_y,n_z$ integers, restricts it to $i \neq
j$ for ${\bf n}=0$. The parameter $\eta$ governs the convergence of the
real-space and reciprocal contributions to the energy. With $\eta = 6.5/L$, adopted in our 
calculations, only terms with ${\bf n}=0$ need to be retained in   
Eq.\ (\ref{I2}). The sum in reciprocal space extends
over all lattice vectors ${\bf k} = 2 \pi  {\bf n}/L$  with  $|{\bf
  n}^2| \le 36$.

\subsection{\label{GC-free}The hypersphere method for Yukawa interactions.}

In the $\mathcal{S}_{3}$ geometry the Green's function of
the Helmholtz equation for the Yukawa fluid is obtained analytically
\cite{Cai-Gilles,Cai-Gilles2}:
\begin{eqnarray}
v^{\mathcal{S}_{3}}(\psi) &=& \frac{1}{R} \frac{\sinh(\omega(\pi-\psi))}
{\sin \psi \sinh(\omega \psi)} \nonumber \; \; \mathrm{for} \; \; \alpha R \geq 1 \; , \\
&=& \frac{1}{R} \frac{\sin(\omega(\pi-\psi))}
{\sin \psi \sin(\omega \psi)} \nonumber \; \; \mathrm{for} \; \; \alpha R \leq 1 \; ,
\end{eqnarray}
where $\omega = (\vert \alpha^2 R^2 -1 \vert)^{1/2}$. 
The pair potential
$v=- q^2 \alpha^{*2} v^{\mathcal{S}_{3}}(\psi)$  in $\mathcal{S}_{3}$ is
isotropic and depends only on the
geodesic length $R\psi$. The geodesic distance between two points
$\bf{R}_1$ and  $\bf{R}_2$ of $\mathcal{S}_{3}$ is given by
\begin{equation}
d_{12} = R \psi_{12} = R \arccos \Big(\frac{\bf{R}_1 \cdot
  \bf{R}_2}{R^2} \Big) \;.
\end{equation}
The configurational energy of a system of $N$ Yukawa hard spheres in $\mathcal{S}_{3}$ is therefore given by
\begin{equation}\label{confi}
V(1, \ldots, N)= \sum_{i<j}v(\psi_{ij})
+ \sum_{i<j}v^{\mathcal{S}_{3}}_{\mathrm{HS}}(\psi_{ij}) - \frac{q^2
  \alpha^{*2}}{2} N V_0 \; .
\end{equation}
In equation\ (\ref{confi}) $v^{\mathcal{S}_{3}}_{\mathrm{HS}}(\psi_{ij})$ denotes the hard sphere potential in $\mathcal{S}_{3}$, i.e.
\begin{eqnarray}
v^{\mathcal{S}_{3}}_{\mathrm{HS}}(\psi) &=0& \; \; R\psi > \sigma \nonumber \\
&=\infty & \; \; R\psi < \sigma \; .
\end{eqnarray}
The presence of the constant $V_0$ in the r.h.s. of equation\
(\ref{confi}) accounts for the fact that the energy is defined up to an
additive constant. The choice of $V_0$ is a recurrent problem of
simulations in $\mathcal{S}_{3}$ (see e.g. the discussion in
reference\cite{Cai-Gilles}). Here we define $V_0$ by requiring that the
self energies in $\mathcal{S}_{3}$  and the usual Euclidean space $\mathbb{R}^3$
coincide, i.e., $V_0=\lim_{r\to 0} \left(  \exp(-\alpha r)/r
-v^{\mathcal{S}_{3}}(r/R) \right) $ which gives 
\begin{eqnarray}
V_0&=& -\alpha + \frac{\omega}{R }\cot(\omega \pi) \; \; \;\;
\alpha R <1\nonumber \\
   &=& -\alpha + \frac{\omega}{R }\coth(\omega \pi) \; \; \;\;
\alpha R >1\nonumber \\
\end{eqnarray}
Recent Monte Carlo simulations of Yukawa plasmas have been performed with this method\cite{Cai-Gilles2,Cai-Gilles3}. Here we apply it to a liquid of attractive Yukawa spheres.
\section{\label{Th}Theory}
\subsection{\label{SCOZA}SCOZA}
The self-consistent Ornstein-Zernike approximation  in its most
general formulation is a liquid-state theory that introduces in the
relation between the direct correlation function $c(r)$ and the pair
potential $v(r)$ one or more state-dependent parameters that are
determined by imposing consistency between the different routes to
thermodynamics leading to a partial differential equation (PDE) for
the unknown parameter(s). The SCOZA originally grew out of the
semi-analytic solution of the mean spherical approximation (MSA) for
HCY potentials~\cite{scoza:12,scoza:11,scoza:9} and it is
this SCOZA formulation that is considered in the present work. It is
based on the Ornstein-Zernike equation supplemented with an MSA-type
closure relation
\begin{equation}
  \label{eq:SCOZAHS_closure}
\begin{array}{rll}
      g(r)&=0&\mbox{for}\quad r\le\sigma\\
      c(r)&=c_{\tiny\mbox{HS}}(r)+K(\rho,\beta)v(r)&\mbox{for}\quad r>\sigma.
\end{array}      
\end{equation}
Here 
$g(r)$ is the pair distribution function, $c_{\tiny\mbox{HS}}(r)$ is
the direct correlation function of the hard-core reference system
given e.g. in the Waisman parameterisation~\cite{Waisma:73} and
$K(\rho,\beta)$ is the state-dependent parameter, that is not given in
advance but is determined so that thermodynamic consistency is
obtained between the energy and compressibility routes ($\beta
=1/k_{\rm B}T$). The consistency 
requirement leads to a PDE of the form
\begin{equation}                                  \label{consist}
\frac{\partial}{\partial \beta} \left( \frac{1}{\chi_{\rm red}} \right)  
= \rho \frac{\partial^2 u}{\partial \rho^2} \,, 
\end{equation} 
where $\chi_{\rm red}=\left(1 - \rho \* \int c(r)\*d^3r\right)^{-1}$
is the reduced (with respect to the ideal gas) dimensionless
isothermal compressibility given by the compressibility route and $u$
is the excess (over ideal gas) internal energy per volume provided by
the energy route, i.e. $u = 2 \pi \rho^2 \int_{\sigma} ^{\infty} dr
r^2 v(r) g(r)$. Once $1/\chi_{\rm red}$ is expressed in terms of
the excess internal energy the PDE~eq.(\ref{consist}) can be
transformed into a PDE for $u$
\begin{equation}                                  \label{consist_u}
B(\rho, u) \frac{\partial u}{\partial \beta}   
=   \rho  \frac{\partial^2 u}{\partial \rho^2} .
\end{equation} 
For the general case of an arbitrary attractive tail of the pair
interaction the determination of $B(\rho,u)$ must be done fully
numerically~\cite{scoza:26} and requires an enormous amount of
computational cost. However, this procedure is not necessary for the
fluid considered here: for certain pair potentials, like the hard-core
Yukawa potential or hard-core potentials with linear combinations of
Yukawa and exponential tails - so called Sogami-Ise fluids, the coefficient $B(\rho,u)$ can be
obtained semi-analytically due to the availability of the
semi-analytic MSA solution for these potentials.  Details of the
determination of $B(\rho,u)$ can be found in
Ref.~\cite{scoza:6}, \cite{KSPS}
and~\cite{scoza:28}. While the former formulation of SCOZA in
Ref.~\cite{scoza:6} is based on the Laplace transformation route
for solving the underlying MSA for hard-core Yukawa fluids, we have
used in the present work the reformulation described in
Refs.~\cite{KSPS} and~\cite{scoza:28} which is based on the
Wertheim-Baxter factorisation method for the solution of the MSA.
While both solution methods are equivalent for the potential
considered here, the Wertheim-Baxter factorisation method is more
flexible and allows a broader applicability of the
SCOZA ~\cite{SPKEPL,SPKJCP}.

The PDE is a quasilinear diffusion equation that has been solved
numerically via an implicit finite-difference algorithm  in the region
$(\beta^\star,\rho^\star)\in[0,\beta_{f}]\times[0,1]$ with suitable initial and
boundary conditions (see Ref.~\cite{scoza:8} for more details). We have chosen a
density and temperature grid spacing of $\Delta \rho^\star= 10^{-4}$ and
$\Delta \beta^\star=2\cdot 10^{-4}$.
 Up to now SCOZA has
been applied to a number of discrete~\cite{scoza:1,scoza:8} and
continuum systems~\cite{scoza:6,plwk:bin:symm} and the results showed
- when compared with computer simulations - that the theory yields
very accurate predictions for the thermodynamic and structural
properties throughout the temperature-density plane, even near the
phase coexistence and in the critical region. Indeed, SCOZA is one of
very few liquid state theories that do not show serious problems in
the critical region and even exhibit some form of scaling with
non-classical, partly Ising-like critical exponents \cite{scoza:25}.

\subsection{\label{HRT} HRT}

In HRT the pair
potential  $v(r)$ is divided into a repulsive reference potential part
$v_{R}(r)$  and a longer-ranged attractive part $w(r)$. 
In the present case the reference  system is the
hard-sphere fluid, whose thermodynamics and correlations are accurately
described by the Carnahan-Starling equation of state~\cite{Carnah:69} 
and the Verlet-Weis
parameterisation of the two-body radial distribution function~\cite{verlet}. 
The HRT differs from the conventional liquid state approaches
in the way the attractive perturbation is dealt with:
in order to take into account the effect of fluctuations,
$w(r)$
 is introduced  gradually via a sequence of $Q$ systems 
whose interaction potential $v_Q(r)$
results from the sum of $v_R(r)$ and a perturbation $w_Q(r)$.
The latter term is defined by its Fourier
transform $\widetilde{w}_Q(k)$ which coincides with 
$\widetilde{w}(k)$ for $k>Q$, and is identically vanishing for $k<Q$. 
As $Q$
evolves from $Q=\infty$ to $Q=0$, the interaction $v_{Q}(r)=v_{R}(r)+w_{Q}(r)$ 
goes from the reference part $v_{R}(r)$ to the full potential $v(r)$. 
In other words the parameter $Q$ plays the role of an infrared cutoff whose  effect consists in depressing fluctuations on length scales larger than $1/Q$:
the critical fluctuations 
are recovered only in the limit $Q\rightarrow 0$. The corresponding 
evolution of 
thermodynamics and correlations of increasing order
can be determined via perturbation theory 
and is described by an exact hierarchy of integro-differential equations.
As mentioned above
close to a critical point 
and at large length scales, this hierarchy becomes  equivalent to 
a formulation of the momentum-space renormalisation group~\cite{nicoll}. 
Unlike current liquid-state theories
HRT preserves the correct convexity of the free energy also below the critical temperature,  producing flat isotherms in the
coexistence region which correspond to infinite compressibility and constant chemical potential. 
The first equation of the hierarchy gives the evolution of the Helmholtz free
energy $A_{Q}$ of the partially interacting system in terms of its two-body 
direct correlation function in momentum space $c_{Q}(k)$ 
and the full perturbation $\widetilde{w}(k)$:
\begin{equation}
\frac{\partial {\cal A}_{Q}}{\partial Q}=-\frac{Q^2}{4\pi^2} \,
\ln \! \left(1-\frac{\Phi(Q)}{{\cal C}_{Q}(Q)}\right) \, 
\label{hrteq}
\end{equation}
where $\Phi(k)=-\widetilde{w}(k)/k_{\rm B}T$. 
The quantities ${\cal A}_{Q}$, ${\cal C}_{Q}(k)$ are linked to $A_{Q}$
and $c_{Q}(k)$ by the relations
\begin{eqnarray}
& & {\cal A}_{Q} = -\frac{A_{Q}}{k_{\rm B}T V}+\frac{1}{2} \rho^{2}
\left[\Phi(k\!=\!0)-\Phi_{Q}(k\!=\!0)\right]
-\frac{1}{2} \,
\rho\!\int\!\!\frac{d^{3}{\bf k}}{(2\pi)^{3}}
\left[\Phi(k)-\Phi_{Q}(k)\right]
\label{amod} \\
& & {\cal C}_{Q}(k) = c_{Q}(k)+\Phi(k)-\Phi_{Q}(k) \, ,
\label{cmod}
\end{eqnarray}
where $V$ is the volume and $\rho$ the number density. These modified free
energy and direct correlation function have been introduced in order to remove
the discontinuities which appear in $A_{Q}$ and $c_{Q}(k)$ at $Q=0$ and $k=Q$
respectively 
as 
a consequence of 
the discontinuity of $\widetilde{w}_{Q}(k)$ 
at $k=Q$. 
Indeed, they represent the free energy 
and direct 
correlation function of the fully interacting fluid as given by a treatment
such that the Fourier components of the interaction with wavelengths larger 
than $1/Q$ are not  really disregarded, but 
approximately taken into account
by mean-field theory. 
In the $Q\rightarrow 0$ limit  the modified 
quantities coincide with the physical ones, once the fluctuations have been 
fully included.
For $Q=\infty$ the definitions in Eqs. ~(\ref{amod}) and (\ref{cmod}) 
reduce, instead, to the well-known  mean-field expressions 
of the free energy and direct correlation function. 

A simple approximation scheme consists in truncating the hierarchy at the first equation, supplementing it with a suitable closure relation based on some ansatz for ${\cal C}_{Q}(k)$.
As in previous 
applications, our form of ${\cal C}_{Q}(k)$  has been inspired by  
liquid-state theories, 
\begin{equation}
{\cal C}_{Q}(k)=\widetilde{c}_{\rm HS}(k)+\lambda_{Q}\,\Phi(k)+{\cal G}_{Q}(k)\, ,
\label{closure}
\end{equation}
where $\widetilde{c}_{\rm HS}(k)$ is the Fourier transform of the direct correlation function of the hard-sphere
fluid, and $\lambda_{Q}$, ${\cal G}_{Q}(k)$ are {\it a priori} unknown 
functions of the thermodynamic state  and of Q.  
The function 
${\cal G}_{Q}(k)$ is determined by the core condition, i.e., the requirement
that the radial distribution function $g_{Q}(r)$ is vanishing for every $Q$ 
whenever the interparticle separation is less than the hard-sphere diameter
$\sigma$.
Via Eq. ~(\ref{closure}) we can write
the core condition
in terms of an integral equation for $c_Q(k)$.
$\lambda_{Q}$ is adjusted so that ${\cal C}_{Q}(k)$ satisfies
the compressibility rule. 
For each Q-system this constraint gives the reduced compressibility
of the fluid as the structure factor evaluated at zero wavevector, and can 
be expressed in terms of the modified quantities ${\cal A}_{Q}$, 
${\cal C}_{Q}(k)$ as
\begin{equation}
{\cal C}_{Q}(k\!=\!0)=\frac{\partial^{2}{\cal A}_{Q}}{\partial \rho^{2}} \, .
\label{sum}
\end{equation}
This equation can be regarded as a consistency condition between the compressibility route and a route in which the thermodynamics is determined from the Helmholtz free energy as obtained from ~\eq{hrteq}.
Here the compressibility rule~(\ref{sum}) plays a fundamental role. 
In fact, when $\lambda_{Q}$ in ~\eq{closure} is determined via  
~\eq{sum} and the resulting expression for ${\cal C}_{Q}(k)$ is used
in ~\eq{hrteq}, one obtains a partial differential equation for 
${\cal A}_{Q}$ which reads  
\begin{equation}
\frac{\partial {\cal A}_{Q}}{\partial Q}=-\frac{Q^2}{4\pi^2} \,
\ln \! \left[1-\frac{\Phi(Q)}{{\cal A}''_{Q} \varphi(Q) + \psi(Q)}\right] \, ,
\label{hrtclos}
\end{equation}
where we have set ${\cal A}''_{Q}=\partial^{2}{\cal A}_{Q}/\partial\rho^{2}$, 
$\varphi(k)=\Phi(k)/\Phi(0)$ and
\begin{equation}
\psi(k)=\widetilde{c}_{\rm HS}(k)+{\cal G}_{Q}(k)-[\widetilde{c}_{\rm HS}(0)+{\cal G}_{Q}(0)]
\varphi(k) \, .
\label{psi}
\end{equation}
\eq{hrtclos} is integrated numerically from $Q=\infty$ down to $Q=0$. 
At each integration step, ${\cal G}_{Q}(k)$ is determined by 
the core condition $g_{Q}(r)=0$, $0<r<\sigma$. This condition
acts as 
an auxiliary equation for determining $\psi(k)$ via \eq{psi}. 

For this specific work  we considered the density interval $\rho^*\in [0,1]$
to numerically solve the differential equation 
for the free energy.
The spacing of the density grid is $\Delta\rho^*=10^{-3}$, 
consequently the error in the determination of the
coexistence  densities 
is of the same order of magnitude.
Concerning the boundary conditions:
 for low density
the system behaves as  an ideal gas
while at 
$\rho^*=1$ we  assumed  the validity of the standard ORPA approximation.

Two remarks are worth mentioning: 
First, Eq. ~(\ref{closure}) assumes that the fluid direct correlation 
function  depends linearly on the perturbation  $\Phi(k)$. 
This ansatz is especially appropriate when the perturbation range is much longer 
than that of the reference part of the interaction, i.e. longer than  the particle size. 
The second remark relates to the implementation of the core condition: 
the inverse Fourier transform of the function ${\cal G}_{Q}(k)$ has been
expanded in a series of Legendre polynomials for $ r <\sigma$ and the series has been truncated after a finite number of terms (typically five). 
The evolution equations for the expansion coefficients 
were then determined by differentiating 
with respect to $Q$
the integral equation 
for $c_Q(k)$,
i.e. the core condition, 
and subsequently projecting it on the polynomials used in the
expansion. However, the resulting equations are coupled to the evolution
equation ~\ref{hrteq} for the free energy, and this makes them difficult
to handle. Therefore, as described in detail in [21, 22], in the
derivative of $c_Q(k)$ with respect to Q the long-wavelength
contributions containing the isothermal compressibility of the Q-system
were disregarded. Physically this amounts to a decoupling of the short- and
the long-range evolutions of the correlations.  
This approximation, as well as the previous one,
 are fully justified for long-range perturbations, where the short- and
 the long-range parts of the correlations are mainly affected by the
 reference and the perturbation terms respectively, but become more
 problematic for short-range interactions where the interplay between
 excluded-volume and cohesion effects is much stronger. However, this effect
 does not concern the present study.

\subsection{\label{OMF} Optimised mean field theory}

We consider, quite generally,  a fluid  of identical hard spheres  of
diameter $\sigma$ with additional isotropic pair interaction
$v(r_{ij})$. Since $v(r)$ is an arbitrary function of $r$ 
for $r \leq \sigma$, one can assume that $v(r)$ has been
regularised in the core in such a way that  its Fourier transform
$\widetilde{v}(k)$ is a well behaved function of $k$ and that $v(0)$ is
a finite quantity.  

At equilibrium at inverse temperature 
$\beta$ and chemical potential  $\mu$ (grand canonical (GC) ensemble)
the pressure $p$ of the fluid is a convex function of $\nu=\beta \mu$
(at fixed $\beta$) 
\cite{Goldenfeld,Cai-dft} even before the thermodynamic limit has been
taken \cite{Goldenfeld,Cai-dft}. As a consequence 
$\beta p(\nu)$ is continuous, its derivatives with respect to $\nu$
exist and it admits a Legendre transform $\beta f \left(\rho \right)$
with respect to $\nu$ defined as 
\begin{eqnarray}
\label{A-leg}
\beta f \left( \rho\right) &=& \sup_{\nu}\left\lbrace 
\rho \nu  -\beta p\left(\nu \right)  \right\rbrace  \; .
\end{eqnarray}
The GC free energy  $\beta f \left(\rho \right)$ is then a convex function of the density and its Legendre transform with respect to $\rho$ coincides with $\beta p(\nu)$.

It was shown in ref.\cite{Cai-Mol} that in the case of an attractive interaction (i.e.
$\widetilde{w}(k) =-\beta\widetilde{v}(k) >0$ for all $q$) we have the
inequality 
\begin{eqnarray}\label{boundf}
\beta f(\beta, \rho) &\leq&  \beta f_{\mathrm{MF}}(\beta, \rho) \nonumber \\
\beta f_{\mathrm{MF}}(\beta, \rho) &=& \beta f_{\mathrm{HS}}(\rho) -\frac{1}{2} \rho^2
\widetilde{w}(0) + \frac{1}{2} \rho w(0) \; .
\end{eqnarray}
Here the subscript mean field (MF) emphasises that $ f_{\mathrm{MF}}$ is
the van der Waals free energy\cite{Hansen}. 
Note that $\beta f_{\mathrm{MF}}\left[ \rho \right]$ is \textit{a priori}  non convex, notably  in the two-phase region.

>From here on we specialise to the Yukawa interaction $w(r)=\beta q^2
\alpha^{*2} y(r) $ with
$y(r)=\exp(-\alpha r)/r$. Since $y(0)=\infty$, equation\ (\ref{boundf})
does not appear to be very useful. However, it follows from appendix A that we can
replace $y(r)$ by the regular function $W_{\tau}(r)/Q_{\tau}^2$ where
$W_{\tau}(r)$ is the interaction energy of two spherical distributions
$\tau(\bf r)$
of Yukawa charges of effective charge $Q_{\tau}$. The interaction is
regularised in the core but remains unchanged for $r>\sigma$. Since
$\widetilde{W}_{\tau}(k) \geq 0$, Eq.\ (\ref{boundf}) is still valid
and will give a non trivial upper bound for the free energy. Similar to
what has been done  in reference\cite{Cai-Coul}   for the
Coulomb interaction we seek a best
upper bound for $\beta f(\beta, \rho)$ by minimising the quantity
\begin{equation}
X = \frac{W_{\tau}(0) - \rho \widetilde{W}_{\tau}(0)}{Q_{\tau}^2} \; .
\end{equation}
To this end we consider variations of $X$ with respect to the charge
distribution $\tau(r)$ defined by Eq.\ (\ref{tau}) and work
out the stationary condition ($\overline{\sigma} = \sigma/2$)
\begin{equation}\label{statio}
\frac{\delta X}{\delta \tau(\bf{r})} =0 \; \; \mathrm{ for} \; \; r <\overline{\sigma} \; .
\end{equation}
First, it follows from equation\ (\ref{Q}) that 
\begin{equation}\label{eq1}
\frac{\delta Q_{\tau}}{\delta \tau(\bf{r})}= \frac{\sinh(\alpha r )}{\alpha r }  \; .
\end{equation}
Second, as a consequence of the convolution relations $V_{\tau}=\tau*y$ and
$W_{\tau}=\tau* V_{\tau}= \tau*y*\tau$ one has
\begin{equation}
\frac{\delta W_{\tau}(\bf{r}')}{\delta \tau(\bf{r})}=2 V_{\tau}(\vert \bf{r}'- \bf{r}\vert ) \; ,
\end{equation}
and thus
\begin{equation}\label{eq2}
\frac{\delta W_{\tau}(0)}{\delta \tau(\bf{r})}=2 V_{\tau}(r) \; .
\end{equation}
Finally, since $\widetilde{W}_{\tau}(0)=4 \pi
\alpha^{-2}\widetilde{\tau}^2(0)$, we have 
\begin{equation}\label{eq3}
\frac{\delta \widetilde{W}_{\tau}(0)}{\delta \tau(\bf{r})}=
\frac{8\pi \widetilde{\tau}(0)}{\alpha^2} \; .
\end{equation}
Collecting   results\ (\ref{eq1}),\ (\ref{eq2}) and \ (\ref{eq3}) one gets
\begin{eqnarray}
\frac{\delta X}{\delta \tau(\bf{r})} &=&
- \frac{2}{Q_{\tau}^3}\; \frac{\sinh(\alpha r)}{\alpha r} \;
\left( W_{\tau}(0) - \rho \widetilde{W}_{\tau}(0) \right) \nonumber \\
&+& \frac{1}{Q_{\tau}^2}\left(
 2 V_{\tau}(r) - \frac{8 \pi \rho \widetilde{\tau}(0)}{\alpha^2}
\right) \; \; \mathrm{ for} \; \; r  <\overline{\sigma} \; .
\end{eqnarray}
Note that we can  impose $\widetilde{\tau}(0)=1$ since, if 
$\tau(r)$ is solution of the stationary condition\ (\ref{statio}), then
$\lambda \tau(r)$, where $\lambda \neq 0$ is an arbitrary constant, is
also a solution ($Q_{\tau}$ is then multiplied by $\lambda$ but
$W_{\tau}/Q_{\tau}^2$ is left unchanged). Therefore Eq.\
(\ref{statio}) can be recast in the form

\begin{eqnarray}\label{resuV}
V_{\tau}(r)&=& \frac{4 \pi \rho}{\alpha^2} + \frac{1}{Q_{\tau}}
+\frac{\sinh(\alpha r)}{\alpha r}\left( 
W_{\tau}(0)-\rho \widetilde{W}_{\tau}(0)
\right).
\end{eqnarray}
Of course, Eq.\ (\ref{resuV}) is valid only for
$r<\overline{\sigma}$. The first contribution in the r.h.s of
Eq.\ (\ref{resuV}) is clearly identified with the potential created by a
uniform
density of Yukawa charges of density $\rho$ while the second
contribution  stems from a charge discontinuity at
$r=\overline{\sigma}$. The solution is therefore of the form  $\tau(r)=\rho
\Theta(\overline{\sigma}) + K \delta(r-\overline{\sigma})$, where $K$ is
a constant determined by the condition $\widetilde{\tau}(0)=1$,
yielding

\begin{equation}
\tau(r)=\rho \Theta(\overline{\sigma}) + \frac{1- \eta}{\pi \sigma^2}\delta(r-\overline{\sigma}) \;
\end{equation}
($\eta=\pi \rho \sigma^3/6$ packing fraction). Furthermore, 
from Eq.\ (\ref{Q}) it follows that
\begin{equation}
Q_{\tau}= 3 \eta \frac{\alpha\overline{\sigma} \cosh(\alpha\overline{\sigma})-
\sinh(\alpha\overline{\sigma}) }{(\alpha\overline{\sigma})^3}
- (1-\eta) \frac{\sinh(\alpha\overline{\sigma})}{\alpha\overline{\sigma}} \; 
\end{equation}
so that the potential $V_{\tau}$  is seen to be   the superposition of
two potentials created by spherical surface and  volume 
distributions, the expressions of which are given in
appendix A by Eqs.\ (\ref{surf}) and\ (\ref{vol}), respectively. 
One finds
\begin{eqnarray}
V_{\tau}(r) &=& \frac{24 \eta}{\alpha^{* 2}} -
\frac{2e^{-\alpha^{*}/2}}{\alpha^{* 2}} \; 
\left(  \eta \alpha^{* 2} + 12 \eta + 6\eta \alpha^{*} -\alpha^{* 2}
\right) \frac{\sinh(\alpha r)}{\alpha r} \; .
\end{eqnarray}
One can check that $V_{\tau}$ indeed verifies Eq.\
(\ref{resuV}).
Finally, the optimised upper bound for the free energy can then be shown to be equal to
\begin{eqnarray}
\beta f(\beta, \rho) &\leq& \beta f_{\mathrm{OMF}}(\beta,\rho) \equiv \beta f_{\mathrm{HS}}(\rho) + \Delta_{\mathrm{OMF}}(\beta, \rho)   \nonumber \\
\Delta_{\mathrm{OMF}} &=& 
\frac{ 6 \beta \eta \alpha^{* 3} e^{-\alpha^{*}} }{\pi} \times \nonumber \\
&\times& \frac{
\eta \alpha^{* 2} + 6 \eta \alpha^{*} + 12 \eta -\alpha^{* 2} 
}{(12 \eta - 6 \eta\alpha^{*} - \alpha^{* 2} +\alpha^{* 2}\eta ) -
  e^{-\alpha^{*}} (\alpha^{* 2} - \alpha^{* 2}\eta -6 \eta \alpha^{*}
  -12 \eta)} \\ 
&=& \frac{-72 \eta^2}{\pi} - \frac{6 \eta (-5 -5 \eta + \eta^2)}{5 \pi}
\alpha^{*2}  + O(\alpha^{*3}) \; .
\end{eqnarray}

An optimised mean field (OMF) theory for the fluid can now be devised by considering the Landau function\cite{Goldenfeld}
\begin{equation}\label{landau}
\mathcal{L}(\beta,\rho,\nu)= \nu \rho - \beta f_{\mathrm{OMF}}(\beta,\rho) \; .
\end{equation}
At given $\beta$ and $\nu$, the density $\rho$ is the one which minimises
the Landau function $\mathcal{L}(\beta,\rho,\nu)$. 
In the limit $\alpha^{*} \to 0$, $\Delta_{\mathrm{OMF}}=-2 \pi \rho^2
\beta$ and one thus revovers the free energy of the Kac model for
$\alpha =0$.

\section{\label{results} Results}
\begin{figure}[p]
\vbox{\noindent\includegraphics[width=\hsize,angle=0]{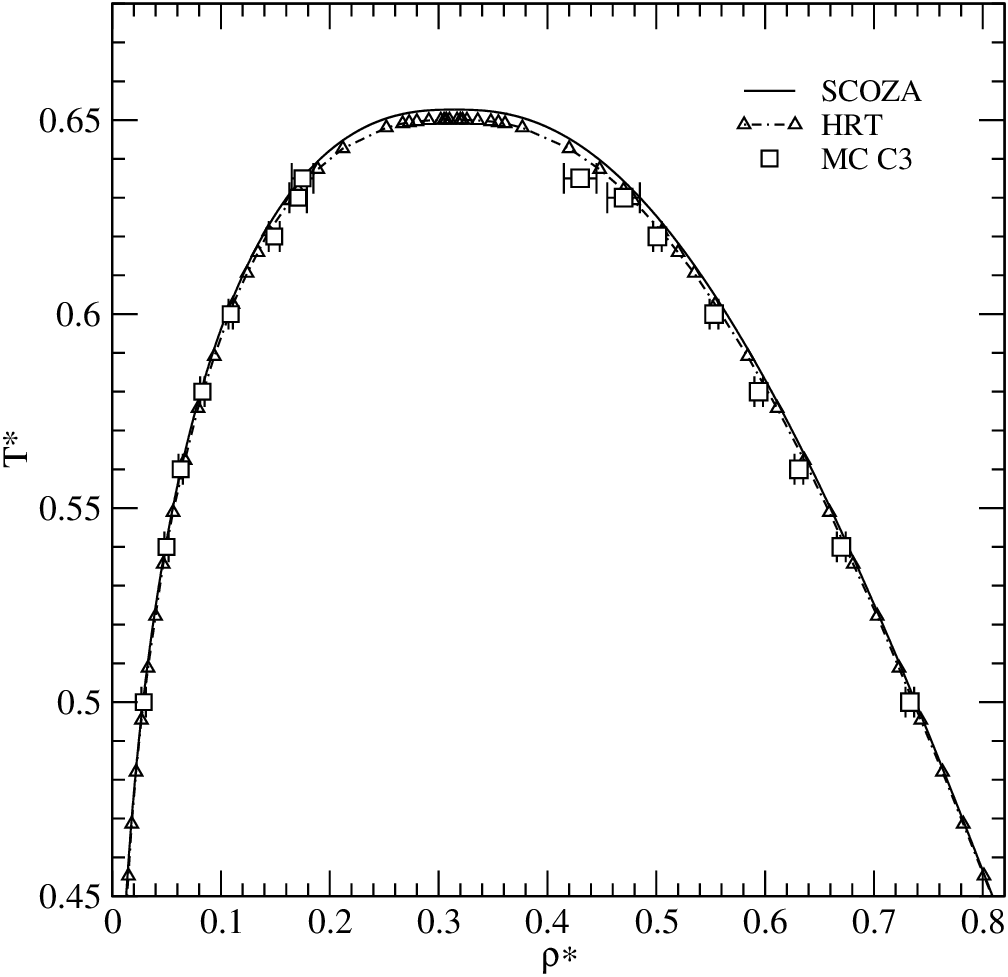} \caption{Liquid-vapour coexistence curve
  of the HCY fluid for $\alpha^*=1.8$ from SCOZA and HRT. The simulation
  results (open squares) are for the cubic system with EW sums ($N=1000$).}}
\label{figure1}
\end{figure}

\begin{figure}[p]
\vbox{\noindent\includegraphics[width=\hsize,angle=0]{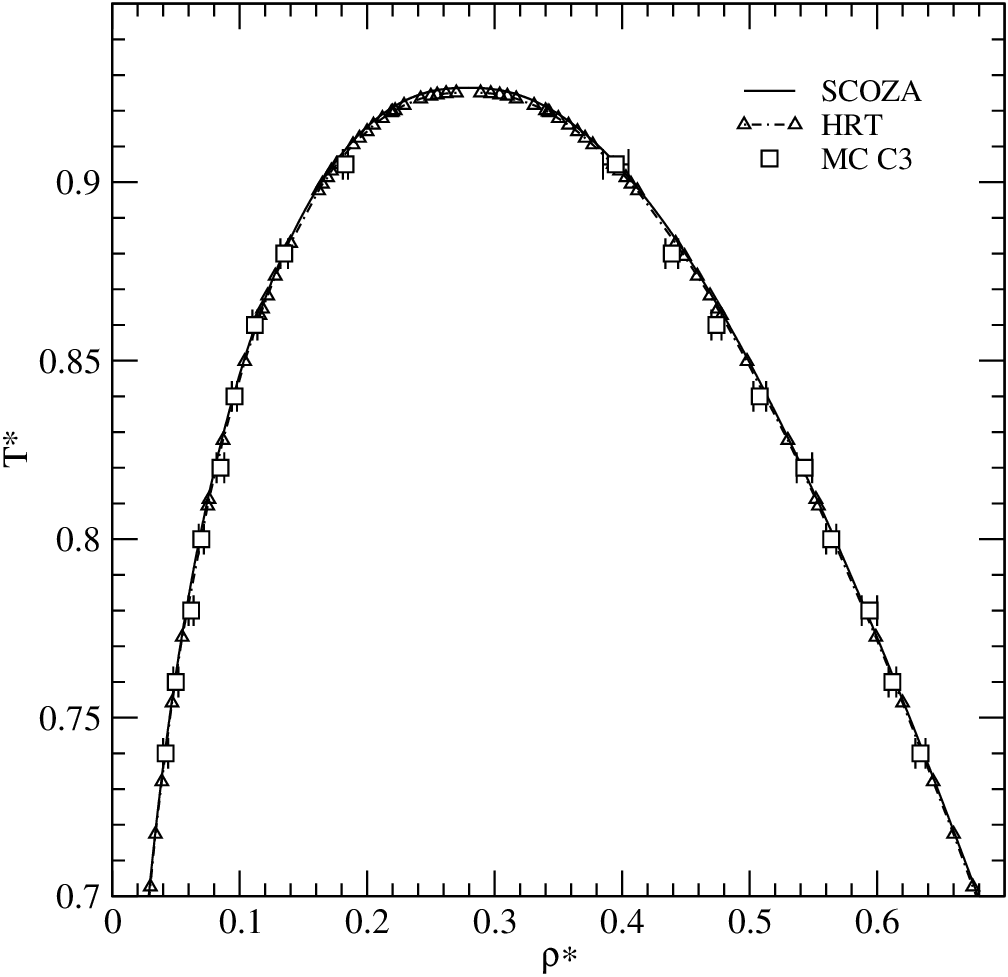} \caption{Liquid-vapour coexistence curve 
  of the HCY fluid for $\alpha^*=1.0$ from SCOZA and HRT. The simulation
 results (open squares) are for the cubic system with EW sums ($N=1000$).}}
\label{figure2}
\end{figure}

\begin{figure}[p]
\vbox{\noindent\includegraphics[width=\hsize,angle=0]{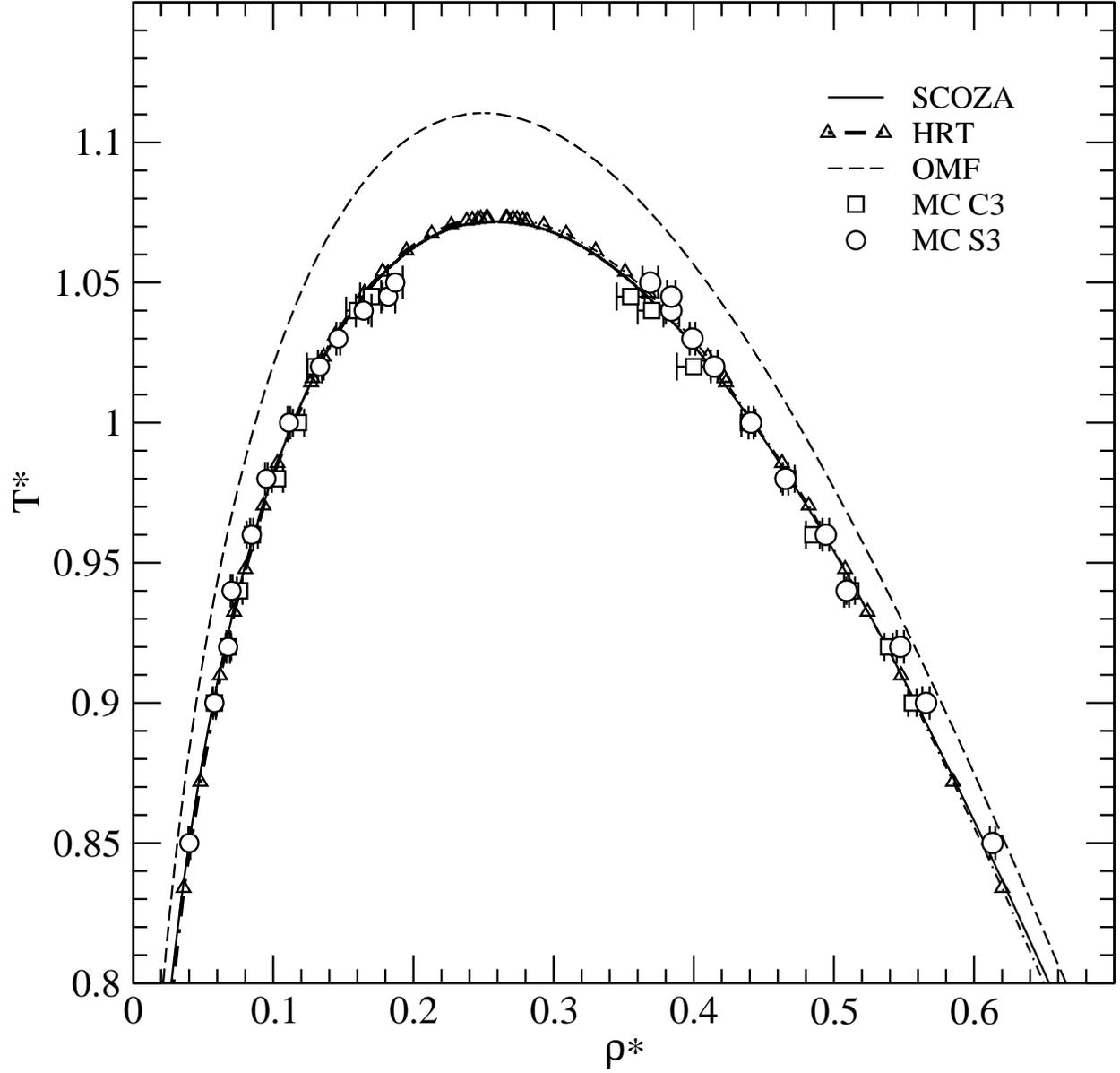} \caption{Liquid-vapour coexistence curve
  of the HCY fluid for $\alpha^*=0.5$ from SCOZA, HRT and OMF. Open
  squares: simulation results  for the cubic system with EW sums
  ($N=1000$); open circles: hypersphere calculations  ($N=2000$).}}             
\label{figure3}
\end{figure}

\begin{figure}[p]
\vbox{\noindent\includegraphics[width=\hsize,angle=0]{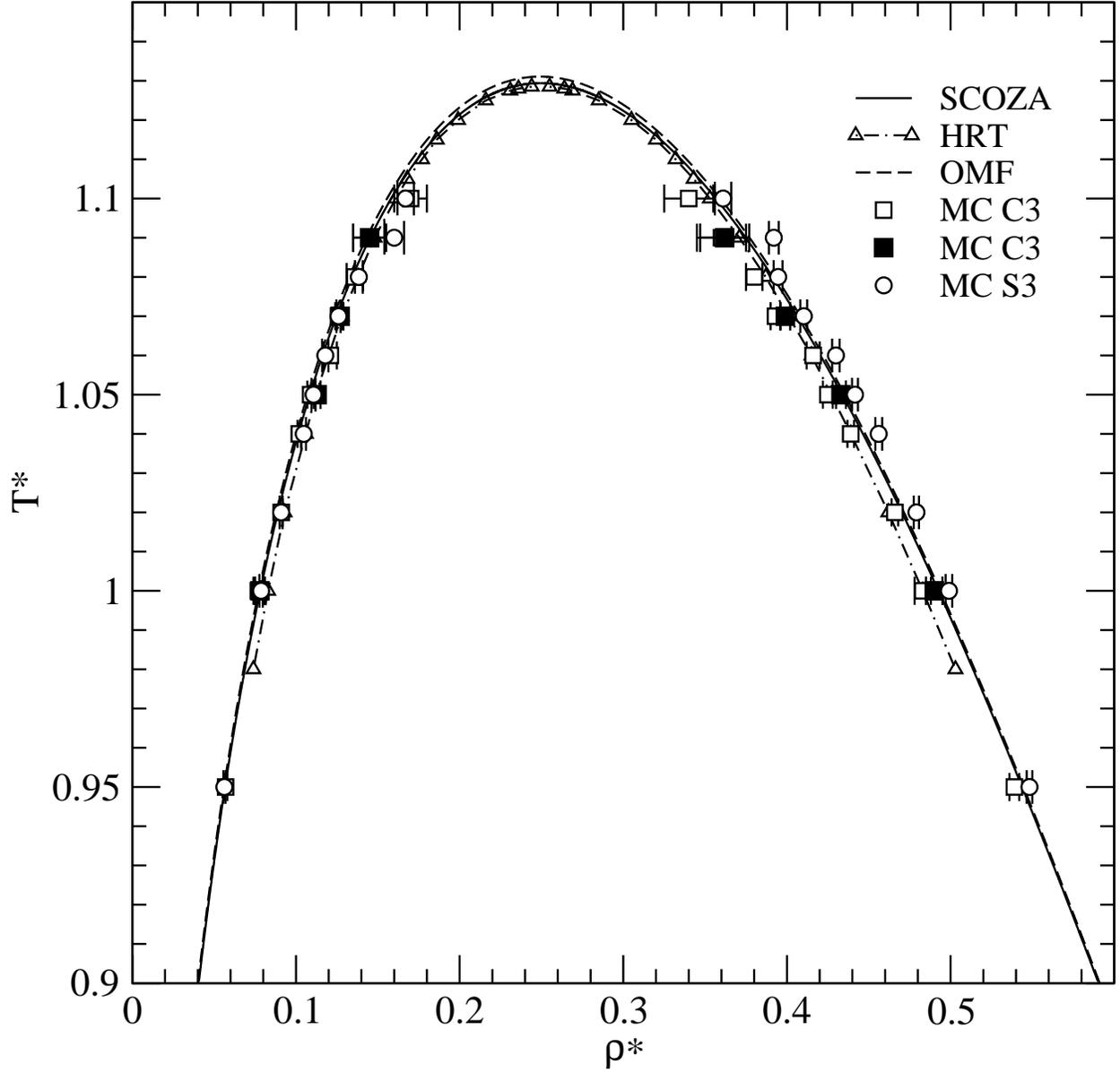} \caption{Liquid-vapour coexistence curve
  of the HCY fluid for $\alpha^*=0.1$ from SCOZA, HRT and
  OMF. Simulation results are for the cubic system with EW sums
  (open squares: $N=1000$, filled squares: $N=1872$); and hypersphere
  calculations  (open circles $N=2000$).}}             
\label{figure4}
\end{figure}

A comparison between the three theoretical approaches  and simulation
data for the liquid-vapour
coexistence curve is made in Figs.\ \ref{figure1}
-\ref{figure4} for $\alpha^*=1.8,1.0,0.5$ and 0.1. 
Values for the densities, internal energies, chemical
potentials and Helmholtz free energies along the coexistence curve are
presented in Tables \ref{table1}-  \ref{table4}. These data may be
valuable for reference to future approaches.

The simulation results have been obtained with the Gibbs ensemble Monte
Carlo (GEMC) method \cite{Panagi:87,Frenkel} using mostly a total of
1000 particles (EW b.c.)  and 2000 particles ($\mathcal{S}_{3}$ b.c.).
The number of cycles generated after equilibration varied from 7-40
$\times 10^6$ according to temperature, each cycle comprising with equal
probabilities,  displacement of
the $N$ particles, volume change, or 
exchange of 500  (EW b.c.) or 1000 particles ($\mathcal{S}_{3}$ b.c.).
Error bars (two standard deviations) were calculated by averaging
densities, energies or histograms over blocks of 1 million cycles.
Expectedly, uncertainties are largest near the critical temperature where
fluctuations of density of the coexisting phases are important and hence
convergence slow. Finite size effects additionally affect the critical
behaviour, but also influenced to some extent the determination of the coexistence
density on the liquid side at least for the smallest value 
$\alpha^*=0.1$.
Thus, for the $\mathcal{S}_{3}$  b.c., an increase of the number of particles from 1000
to 2000 typically lowered $\rho_l^*$ by about $3\%$. The opposite trend was
observed with the EW b.c. where  $\rho_l^*$ increased by $ 1-2 \%$  by an
increase of $N$ from 1000 to 1872 (cf.\ table \ref{table4}).
It is likely that, for $\alpha^*=0.1$, small finite size effects are
still present in both sets of results.

\begin{figure}[p]
\vbox{\noindent\includegraphics[width=\hsize,angle=0]{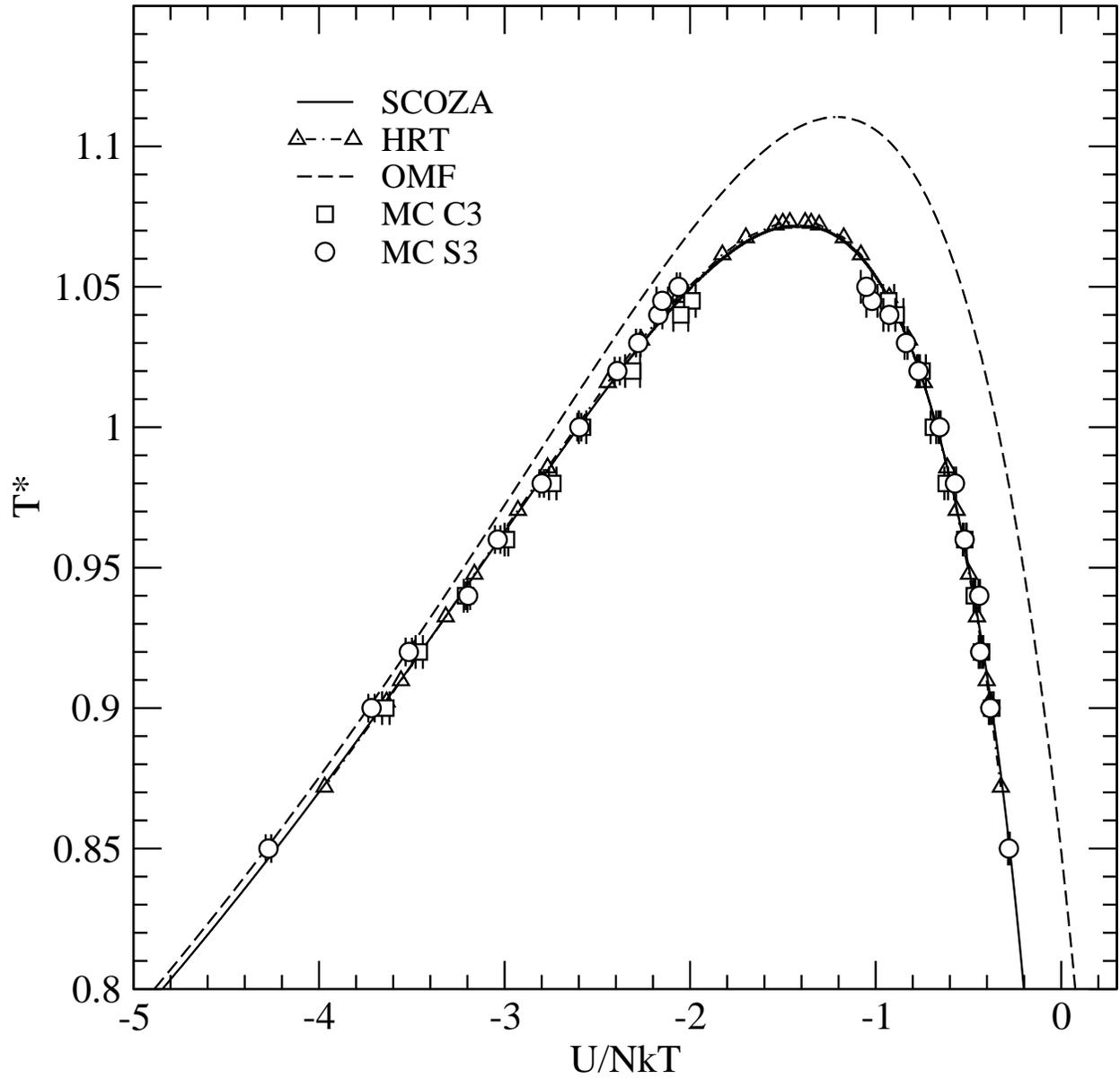} \caption{Internal energy $U/Nk_{\rm B}T$
  along the coexistence curve
  for $\alpha^*=0.5$. The symbols are as in Fig.\ 3.}}
\label{figure5}
\end{figure}

The agreement between HRT, SCOZA and MC simulations turns out to be
extremely good over the range of $\alpha$ values considered. Small
differences occur only at the highest and lowest values of $\alpha$.
At $\alpha^*=1.8$ the critical temperature of HRT is slightly below that
predicted by SCOZA. Previous work \cite{Caccam:99} showed, however, 
 that by further increasing the value of  $\alpha$ the HRT critical
temperature  eventually gets higher than that of SCOZA.
At $\alpha^*=0.1$ the coexistence curves, including the critical
regions, of SCOZA and OMF are very close (cf.\ Fig.\ \ref{figure4}); as the OMF theory is expected
to be nearly exact at this small value of $\alpha$ it is very likely
that SCOZA becomes exact also in the Kac limit, though no formal proof seems
to be available. This is further confirmed by the close agreement
between critical parameters of SCOZA and OMF at  $\alpha^*=10^{-5}$ (see
table \ref{table5}).  
The coexistence curve of HRT is found to be slightly
narrower than that of SCOZA though in the near critical region the
agreement between the two theories is quite good. As pointed out in
ref. \cite{Caillo:06} the HRT flow cannot be solved conveniently
in the Kac limit. 
%
Inspection of Fig.\ \ref{figure3} shows that the OMF
theory rapidly deteriorates with increase of $\alpha$.
Figure \ref{figure5} compares HRT, SCOZA, OMF and MC simulation results
for the internal energy along the coexistence curve at
$\alpha^*=0.5$. The agreement is similar to that for
the densities. 

The variation with  $\alpha$ of the critical temperature and density for
SCOZA and HRT,
including  data of ref.\  \cite{Caccam:99} for  $\alpha^* > 1.8$, is
given in  table \ref{table5}. 
 No attempt has been made to determine
critical data in the simulations by fitting data away from the critical
point by some power law expression. Because of the finite
size effects in the critical region  uncertainties of the results would
be   large 
and of no use to validate the theoretical predictions.
In the GEMC method   
the order parameter $M=\rho_l-\rho_g$  is believed to have MF behaviour
(critical exponent $\beta=1/2$)  in the critical region once the
correlation length exceeds the linear dimension $L$ of the simulation box . In fact we find
that for $\alpha^* \lesssim 1.0$, $M^2$ varies linearly over the whole
temperature region considered yielding an effective exponent
$\beta_{eff}=1/2$). Such an analysis was not conclusive for $\alpha^*=1.8$.
  
\section{\label{conclusion} Conclusion}
We have compared the predictions for the liquid-vapour coexistence curve of a
long-range Yukawa fluid obtained from advanced theoretical approaches with
Gibbs ensemble Monte Carlo simulations. Concerning the simulations care has to
be taken when the range of the potential exceeds the length of the simulation
box. This was done by performing an Ewald sum in the case of a cubic
simulation box and by using hyperspherical boundary conditions. The
theoretical approaches that we applied comprised the self-consistent
Ornstein-Zernike approximation (SCOZA), the hierarchical reference theory
(HRT), as well as an optimised mean field theory (OMF). While the OMF yields
the exact result in the limit of infinite range of the potential, it
deteriorates with decreasing interaction range. On the other hand, HRT and
SCOZA turn out to be in perfect agreement with simulation results over the
whole interaction range considered. This study complements a recent
comparison~\cite{Caccam:99} for the case of intermediate and short range
Yukawa fluid.

\section*{Acknowledgements}
This work was supported by a grant through the Programme
d'Actions Int{\'e}gr{\'e}es AMADEUS under Project Nos.  06648PB and FR
09/2007,  by the Hochschuljubil\"aumsstiftung der Stadt Wien under
Project Number 1757/2007 and by the Marie Curie European Network MRTN-CT-2003-504712.
Federica Lo Verso thanks Davide Pini, Alberto Parola and Luciano Reatto
for helpful discussions.


\newpage
%
\begin{table}
{\tiny Table 1. Coexistence densities, internal energy $U$, excess chemical
  potential $\mu$ and Helmholtz free energy $A$ at coexistence for
  $\alpha^* = 1.8$. $A^{ref}$ is the Carnahan-Starling HS free energy
  \cite{Carnah:69}.\\

{\begin{tabular}{@{}c l l l l l l l l}\toprule
 $T^*$ &    & $\rho^*_v$  & $\rho^*_l$ & ${(U/NkT)}_v$ & $ {(U/NkT)}_l$
 & $ \mu /kT$ &${(\Delta a)}_v^{\rm a}$  & ${(\Delta a)}_l^{\rm b}$
 \\ \colrule  
      &SCOZA&0.174   &0.464  & -1.055   & -2.650  &-2.737 &    &    \\
0.635 &HRT  &  0.181   & 0.458    & -1.106    & -2.607    &     & 0.963   & 2.521       \\
      &MC-EW&0.175(10)&0.430(15)  & -1.06(4)   & -2.43(3)  &-2.74(1)&  &
 \\ 
\multicolumn{9}{c}{   } \\ 
     &SCOZA&0.161   &0.485  & -0.982   & -2.770  &-2.765 &    &    \\
0.63 &HRT  &   0.166  &   0.477  &   -1.030    &  -2.735    &     & 0.890   & 2.657      \\
      &MC-EW&0.171(8)&0.470(15)  & -1.05(2)   & -2.67(2)  &-2.76(1)&  &
 \\ 
\multicolumn{9}{c}{   } \\ 
     &SCOZA&0.138   &0.515  & -0.862   & -2.996  &-2.825 &    &    \\
0.62 &HRT  &   0.142   &  0.509   &  -0.901    &  -2.967    &     &  0.774  &    2.900   \\
      &MC-EW&0.149(5)&0.501(4)  & -0.94(2)   & -2.92(2)  &-2.83(1)&  &
 \\ 
\multicolumn{9}{c}{   } \\ 
     &SCOZA&0.105   &0.565  & -0.686   & -3.407  &-2.953 &    &    \\
0.60 &HRT  &   0.108   &  0.560   &   -0.724   &   -3.386   &     &   0.609  & 3.333        \\
      &MC-EW&0.109(2)&0.553(4)  & -0.73(1)   & -3.34(2)  &-2.96(1)&  &
 \\ 
\multicolumn{9}{c}{   } \\ 
     &SCOZA&0.082   &0.606  & -0.556   & -3.803  &-3.094 &    &    \\
0.58 &HRT  &   0.083   &  0.602   & -0.578     &   -3.787   &     & 0.486    &   3.741   \\
      &MC-EW&0.083(2)&0.594(4)  & -0.58(1)   & -3.74(2)  &-3.10(1)&  &
 \\ 
\multicolumn{9}{c}{   } \\ 
     &SCOZA&0.063   &0.643  & -0.452   & -4.204  &-3.251 &    &    \\
0.56 &HRT  &  0.065    &  0.640   &   -0.490    &   -4.192    &     & 0.396    &   4.155     \\
      &MC-EW&0.063(2)&0.631(4)  & -0.46(1)   & -4.13(3)  &-3.25(3)&  &
 \\ 
\multicolumn{9}{c}{   } \\ 
     &SCOZA&0.049   &0.676  & -0.367   & -4.614 &-3.424 &    &    \\
0.54 &HRT  &    0.050  &  0.674   &  -0.390     & -4.606    &     &  0.317  &4.573       \\
      &MC-EW&0.050(2)&0.670(4)  & -0.39(1)   & -4.57(2)  &-3.47(8)&  &
 \\ 
\multicolumn{9}{c}{   } \\ 
     &SCOZA&0.0284   &0.738  & -0.235   & -5.503  &-3.834 &    &    \\
0.50 &HRT  &   0.029   &   0.736  &   -0.258    &  -5.500    &     &  0.201  &5.469        \\
      &MC-EW&0.029(2)&0.733(4)  & -0.25(1)   & -5.47(3)  &-3.76(10)&  &
 \\ 
\botrule
\end{tabular}}\\
$^{\rm a}$${(\Delta a)}_v=\frac{1}{NkT}{(A-A^{ref})}_v$\\
$^{\rm b}$${(\Delta a)}_l=\frac{1}{NkT}{(A-A^{ref})}_l$
\label{table1}}
\end{table}
\newpage

\begin{table}
{\tiny Table 2. Same as table 1 but for $\alpha^* = 1.0$\\
{\begin{tabular}{@{}c l l l l l l l l}\toprule
 $T^*$ &    & $\rho^*_v$  & $\rho^*_l$ & ${(U/NkT)}_v$ & $ {(U/NkT)}_l$
 & $ \mu /kT$ &${(\Delta a)}_v^{\rm a}$  & ${(\Delta a)}_l^{\rm b}$
 \\ \colrule  
      &SCOZA&0.174   &0.396 & -0.987   & -2.209  &-2.773 &    &    \\
0.905 &HRT  &0.176  &0.393  &-1.000   &-2.195   &     &0.955    &2.169  \\
      &MC-EW&0.183(2)&0.395(10)  & -1.025(50)   & -2.19(5)  &-2.77(1)&  &
 \\ 
\multicolumn{9}{c}{   } \\ 
     &SCOZA&0.135   &0.450  & -0.788   & -2.583  &-2.864 &    &    \\
0.88 &HRT  & 0.136  &0.448  &-0.767      &-2.477 &     &0.758 &2.557  \\
      &MC-EW&0.135(2)&0.439(5)  & -0.78(2)   & -2.52(2)  &-2.87(1)&  &
 \\ 
\multicolumn{9}{c}{   } \\ 
     &SCOZA&0.112   &0.484  & -0.676   & -2.852  &-2.944 &    &    \\
0.86 &HRT  &0.114  &0.482   &-0.687  &-2.842   &     & 0.650   &2.825 \\
      &MC-EW&0.112(2)&0.474(4)  & -0.67(1)   & -2.79(2)  &-2.95(1)&  &
 \\ 
\multicolumn{9}{c}{   } \\ 
     &SCOZA&0.095   &0.514  & -0.586   & -3.107  &-3.028 &    &    \\
0.84 &HRT  &0.096   &0.512  & -0.593   & -3.098  &     &0.561 &3.083  \\
      &MC-EW&0.096(2)&0.508(5)  & -0.59(2)   & -3.07(3)  &-3.04(2)&  &
 \\ 
\multicolumn{9}{c}{   } \\ 
     &SCOZA&0.0806   &0.541  & -0.511   & -3.360  &-3.119 &    &    \\
0.82 &HRT  &0.081   &0.540    &-0.513   &-3.353   &     & 0.485   &3.341  \\
      &MC-EW&0.085(3)&0.543(6)  & -0.54(2)   & -3.37(3)  &-3.11(2)&  &
 \\ 
\multicolumn{9}{c}{   } \\ 
     &SCOZA&0.0685   &0.567 & -0.446   & -3.615  &-3.216 &    &    \\
0.80 &HRT  &0.069   &0.566 &-0.451 & -3.610 &     &0.424    &3.600       \\
      &MC-EW&0.070(2)&0.564(4)  & -0.447(15)   & -3.60(3)  &-3.21(1)&  &
 \\ 
\multicolumn{9}{c}{   } \\ 
     &SCOZA&0.058   &0.591  & -0.389   & -3.873 &-3.321 &    &    \\
0.78 &HRT  &0.059   &0.590 &-0.401  &-3.868 &     &0.372 &3.859    \\
      &MC-EW&0.062(2)&0.594(6)  & -0.420(15)   & -3.89(3)  &-3.30(2)&  &
 \\ 
\multicolumn{9}{c}{   } \\ 
     &SCOZA&0.049   &0.614  & -0.339   & -4.140  &-3.434 &    &    \\
0.76 &HRT  &0.050  &0.613   &-0.353  & -4.133 &     & 0.324 &4.124  \\
      &MC-EW&0.050(2)&0.612(3)  & -0.344(5)   & -4.120(15)  &-3.43(1)&  &
 \\ 
\multicolumn{9}{c}{   } \\ 
     &SCOZA&0.041   &0.636  & -0.294   & -4.413  &-3.557 &    &    \\
0.74 &HRT  &0.042   &0.635  &-0.304    &-4.407   &     &0.280 &4.398   \\
      &MC-EW&0.042(2)&0.634(4)  & -0.30(1)   & -4.440(25)  &-3.55(1)&  &
 \\ 
\botrule
\end{tabular}}\\
$^{\rm a}$${(\Delta a)}_v=\frac{1}{NkT}{(A-A^{ref})}_v$\\
$^{\rm b}$${(\Delta a)}_l=\frac{1}{NkT}{(A-A^{ref})}_l$
\label{table2}}
\end{table}
\newpage

\begin{table}
{\tiny Table 3. Same as table 1 but for $\alpha^* = 0.5$\\
{\begin{tabular}{@{}c l l l l l l l l}\toprule
 $T^*$ &    & $\rho^*_v$  & $\rho^*_l$ & ${(U/NkT)}_v$ & $ {(U/NkT)}_l$
 & $ \mu /kT$ &${(\Delta a)}_v^{\rm a}$  & ${(\Delta a)}_l^{\rm b}$
 \\ \colrule  
      &SCOZA&0.164   &0.367 & -0.923   & -2.063  &-2.824 &    &    \\
1.045 &HRT  &0.163   &0.371 & -0.916   & -2.086  &     & 0.908 &2.082   \\
      &MC-EW&0.169(8)&0.355(8)  & -0.93(3)   & -1.99(2)  &-2.82(1)&  & \\
      &MC-S3&0.182(5)&0.384(3)  & -1.02(3)   & -2.15(3)  &    &  &
 \\ 
\multicolumn{9}{c}{   } \\ 
     &SCOZA&0.157   &0.377  & -0.885   & -2.131  &-2.839 &    &    \\
1.04 &HRT  &0.156   &0.381  & -0.881   &-2.153   &     &0.873 &2.149  \\
     &MC-EW&0.160(8)&0.370(8)  & -0.89(4)   & -2.05(4)  &-2.84(1)&  & \\
     &MC-S3&0.165(6)&0.384(6)  & -0.93(3)   & -2.172(22)  &    &  &
 \\ 
\multicolumn{9}{c}{   } \\ 
     &SCOZA&0.133   &0.412  & -0.764   & -2.377  &-2.899 &    &    \\
1.02 &HRT  & 0.132  &0.415  &-0.760   &-2.394  &     &0.752 &2.390        \\
     &MC-EW&0.130(6)&0.400(6)  & -0.75(2)   & -2.31(4)  &-2.90(1)&  & \\
     &MC-S3&0.133(2)&0.415(3)  & -0.77(1)   & -2.39(2)  &     &  &
 \\ 
\multicolumn{9}{c}{   } \\ 
     &SCOZA&0.114   &0.442  & -0.670   & -2.602  &-2.962 &    &    \\
1.00 &HRT  &0.114 &0.445 &-0.670 & -2.620 &     &0.662  &2.617   \\
     &MC-EW&0.118(4)&0.439(4)  & -0.689(15)   & -2.58(2)  &-2.96(1)&  & \\
     &MC-S3&0.111(1)&0.441(2)  & -0.655(50)   & -2.60(1)  &    &  &
 \\ 
\multicolumn{9}{c}{   } \\ 
     &SCOZA&0.099   &0.469  & -0.594   & -2.817  &-3.030 &    &    \\
0.98 &HRT  &0.099  & 0.470 &-0.594  & -2.826 &     &0.587    &2.824  \\
     &MC-EW&0.103(4)&0.467(4)  & -0.62(1)   & -2.74(2)  &-3.02(1)&  & \\
     &MC-S3&0.095(1)&0.466(2)  & -0.573(54)   & -2.800(15)  &    &  &
 \\ 
\multicolumn{9}{c}{   } \\ 
     &SCOZA&0.086   &0.493 & -0.528   & -3.028  &-3.101 &    &    \\
0.96 &HRT  &0.086   &0.494 &-0.527  &-3.035 &     &0.520 & 3.032      \\
     &MC-EW&0.085(4)&0.485(4)  & -0.52(1)   & -2.99(1)  &-3.11(1)&  & \\
     &MC-S3&0.0846(11)&0.494(3)  & -0.521(7)   & -3.037(15)  &     &  &
 \\ 
\multicolumn{9}{c}{   } \\ 
     &SCOZA&0.075   &0.516  & -0.471   & -3.239 &-3.178 &    &    \\
0.94 &HRT  &0.076     &0.516 &-0.476 &-3.240 &     &0.469 & 3.238  \\
     &MC-EW&0.076(2)&0.512(2)  & -0.468(5)   & -3.21(1)  &-3.17(2)&  & \\
     &MC-S3&0.072(1)&0.509(2)  & -0.442(4)   & -3.20(1)  &    &  &
 \\ 
\multicolumn{9}{c}{   } \\ 
     &SCOZA&0.0656   &0.538  & -0.420   & -3.451  &-3.259 &    &    \\
0.92 &HRT  &0.066 & 0.537    & -0.422  &-3.447 &     & 0.417   &3.446  \\
      &MC-EW&0.068(2)&0.539(3)  & -0.43(1)   & -3.46(2)  &-3.27(3)&  & \\
      &MC-S3&0.0678(13)&0.547(3)  & -0.437(8)   & -3.52(2)  &    &  &
 \\ 
\multicolumn{9}{c}{   } \\ 
     &SCOZA&0.0572   &0.559  & -0.374   & -3.667  &-3.346 &    &    \\
0.90 &HRT  &0.058   & 0.557 &-0.380  & -3.658  &     & 0.374 & 3.656  \\
     &MC-EW&0.058(2)&0.556(3)  & -0.375(10)   & -3.64(2)  &-3.34(1)&  & \\
     &MC-S3&0.058(1)&0.566(3)  & -0.382(7)   & -3.72(2)  &     &  &
 \\ 
\botrule
\end{tabular}}\\
$^{\rm a}$${(\Delta a)}_v=\frac{1}{NkT}{(A-A^{ref})}_v$\\
$^{\rm b}$${(\Delta a)}_l=\frac{1}{NkT}{(A-A^{ref})}_l$
\label{table3}}
\end{table}
\newpage

\begin{table}
{\tiny Table 4. Same as table 1 but for $\alpha^* = 0.1$\\
{\begin{tabular}{@{}c l l l l l l l l}\toprule
 $T^*$ &    & $\rho^*_v$  & $\rho^*_l$ & ${(U/NkT)}_v$ & $ {(U/NkT)}_l$
 & $ \mu /kT$ &${(\Delta a)}_v^{\rm a}$  & ${(\Delta a)}_l^{\rm b}$
 \\ \colrule  
      &SCOZA&0.158   &0.357 & -0.900       & -2.030  &-2.863 &    &    \\
1.10 &HRT  & 0.161   &0.353 &  -0.916      & -2.010   &      &  -0.916   & -2.010          \\
     &MC-EW&0.17(1)&0.340(15) & -0.96(2)   & -1.96(5)  &-2.86(2)&  & \\
     &MC-S3&0.167(5)&0.361(5) & -0.93(3)   & -2.05(2)  &        &  &
 \\ 
\multicolumn{9}{c}{   } \\ 
     &SCOZA&0.145   &0.375  & -0.834   & -2.155  &-2.891 &    &    \\
     &HRT  &0.148   &0.371    & -0.850     &    -2.132  &     &-0.850     & -2.131        \\
1.09 &MC-EW&0.145(10)&0.360(15)  & -0.82(1)   & -2.05(1)  &-2.89(1)&  & \\
     &MC-EW1&0.145(4)&0.362(3)  & -0.82(2)   & -2.07(2)  &-2.895(5)&  & \\
     &MC-S3&0.160(6)&0.392(3)  & -0.92(2)   & -2.25(2)  &       &  &
 \\ 
\multicolumn{9}{c}{   } \\ 
     &SCOZA&0.134   &0.391  & -0.778   & -2.270  &-2.918 &    &    \\
1.08 &HRT  &0.138   &0.387  &  -0.800    &   -2.244   &     & -0.800   & -2.244       \\
     &MC-EW&0.136(5)&0.380(5)  & -0.79(2)   & -2.21(2)  &-2.92(2)&  & \\
     &MC-S3&0.1384(24)&0.395(3)  & -0.8018(15)   & -2.29(2)  &   &  &
 \\ 
\multicolumn{9}{c}{   } \\ 
     &SCOZA&0.124   &0.406  & -0.730   & -2.379  &-2.947 &    &    \\
     &HRT  &0.128  &0.401  &  -0.749     &  -2.347   &     & -0.749   &   -2.347    \\
1.07 &MC-EW&0.126(2)&0.393(3)  & -0.73(1)   & -2.32(2)  &-2.95(1)&  & \\
     &MC-EW1&0.127(2)&0.399(3)  & -0.74(1)   & -2.34(2)  &-2.95(1)&  & \\
     &MC-S3&0.1260(15)&0.4103(21) & -0.737(8) & -2.402(13)  &   &  &
 \\ 
\multicolumn{9}{c}{   } \\ 
     &SCOZA&0.116   &0.420  & -0.686   & -2.484  &-2.977 &    &    \\
1.06 &HRT &0.120   &0.415    &  -0.708    &   -2.452   &     & -0.708   & -2.452       \\
     &MC-EW&0.121(4)&0.416(4)  & -0.716(15)   & -2.46(2)  &-2.97(1)&  & \\
     &MC-S3&0.118(2)&0.423(3)  & -0.697(11)   & -2.544(15)  &      &  &
 \\ 
\multicolumn{9}{c}{   } \\ 
     &SCOZA&0.1082   &0.434 & -0.647   & -2.587  &-3.007 &    &    \\
     &HRT  &0.112  &0.427  & -0.667     & -2.547     &    & -0.667   &  -2.547     \\
1.05 &MC-EW&0.109(2)&0.425(3)  & -0.65(1)   & -2.54(2)  &-3.000(15)&  & \\
     &MC-EW1&0.113(2)&0.433(3)  & -0.68(1)   & -2.59(2)  &-3.00(1)&  & \\
     &MC-S3&0.111(2)&0.442(2)  & -0.660(7)  & -2.63(2)  &         &  &
 \\ 
\multicolumn{9}{c}{   } \\ 
     &SCOZA&0.101   &0.446  & -0.610   & -2.689 &-3.038 &    &    \\
1.04 &HRT  &0.107 &0.439    &  -0.638    & -2.644    &     & -0.638   &   -2.644     \\
     &MC-EW&0.102(1)&0.439(2)  & -0.641(5)   & -2.64(1)  &-3.04(1)&  & \\
     &MC-S3&0.1047(13)&0.456(3)  & -0.630(9)   & -2.75(2)  &      &  &
 \\ 
\multicolumn{9}{c}{   } \\ 
     &SCOZA&0.089   &0.470  & -0.546   & -2.887  &-3.103 &    &    \\
1.02 &HRT  &0.093   &0.462     &   -0.571    &   -2.837   &     & -0.570   & -2.837      \\
      &MC-EW&0.091(1)&0.466(2)  & -0.555(5)   & -2.86(1)  &-3.10(1)&  & \\
      &MC-S3&0.091(1)&0.479(2)  & -0.558(6)   & -2.94(1)  &        &  &
 \\ 
\multicolumn{9}{c}{   } \\ 
     &SCOZA&0.078   &0.493  & -0.489   & -3.085  &-3.173 &    &    \\
     &HRT  &0.083   &0.483  &    -0.519  &    -3.026  &     & -0.519    &  -3.026      \\
1.00 &MC-EW&0.077(3)&0.483(5)  & -0.48(1)   & -3.028(25)  &-3.17(1)&  & \\
     &MC-EW1&0.078(2)&0.487(5) & -0.484(14)   & -3.05(3)  &-3.185(15)& & \\
     &MC-S3&0.079(1)&0.499(2)  & -0.492(6)  & -3.126(13)  &     &  &
 \\ 
\multicolumn{9}{c}{   } \\ 
     &SCOZA&0.0572   &0.559  & -0.374   & -3.667  &-3.346 &    &    \\
0.98 &HRT &0.074 &0.503 &  -0.472    &    -3.215  &     & -0.472    &  -3.215      \\
      &MC-EW&0.057(1)&0.539(3)  & -0.373(6)   & -3.55(2)  &-3.35(3)&  &
 \\ 
\botrule
\end{tabular}}\\
$^{\rm a}$${(\Delta a)}_v=\frac{1}{NkT}{(A-A^{ref})}_v$\\
$^{\rm b}$${(\Delta a)}_l=\frac{1}{NkT}{(A-A^{ref})}_l$
\label{table4}}
\end{table}
\newpage
\begin{table}
{\tiny Table 5. Critical parameters. Results based on the Carnahan-Starling
  hard sphere free
  energy \cite{Carnah:69} for the reference system.\\
{\begin{tabular}{@{}c l l l}\toprule
 $\alpha^*$ &    & $T^*_c$  & $\rho^*_c$ 
 \\ \colrule  
0.0   &OMF&1.13194 & 0.24913    \\
\multicolumn{4}{c}{   } \\ 
$10^{-5}$   &SCOZA&1.1319 & 0.2491    \\
\multicolumn{4}{c}{   } \\ 
0.1   &SCOZA&1.1294 & 0.2495    \\
      &HRT  &    1.129   &  0.250\\
      &OMF  &1.13108 & 0.24915  \\
\multicolumn{4}{c}{   } \\ 
0.5   &SCOZA&1.0718 & 0.2586    \\
      &HRT  &    1.073    & 0.260 \\
\multicolumn{4}{c}{   } \\ 
1.0   &SCOZA&0.9264 & 0.2793    \\
      &HRT  & 0.925      &  0.279\\
\multicolumn{4}{c}{   } \\ 
1.8   &SCOZA&0.6527 & 0.3145    \\
      &HRT  &    0.650   &  0.314 \\
\multicolumn{4}{c}{   } \\ 
$4.0^{\rm a}$   &SCOZA&0.173 & 0.3895    \\
      &HRT  &   0.175    & 0.394  \\
\multicolumn{4}{c}{   } \\ 
$7.0^{\rm a}$   &SCOZA&0.0187 & 0.4575    \\
      &HRT  &   0.019     & 0.424  \\
 \botrule
\end{tabular}}\\

$^{\rm a}$ from ref.~\onlinecite{Caccam:99}. Due to a different definition,
  the temperatures of ref.~\onlinecite{Caccam:99} must be multiplied by the
  factor $ \alpha^{*2}/ e^{\alpha^*}$.
\label{table5}}
\end{table}
\newpage


\newpage
\newpage
\newpage
\newpage
\newpage
\newpage
\appendix
\section{Some properties of Yukawa charge distributions}
The electrostatics of Yukawa charge distributions is similar to, but not identical with the electrostatics of usual Coulomb charge distributions.
In this appendix we extend previous results obtained for special forms
of spherical distributions of Yukawa charges
\cite{Rosenfeld1,Rosenfeld2,Rosenfeld3,Rosenfeld4,Cai-Gilles} to a
general spherical distribution $\tau(r) $ which satisfies: 
\begin{eqnarray}\label{tau}
\tau(r)&=&0  \; \; \mathrm{ for }\; \;  r >  \overline{\sigma} \equiv \sigma/2 \; , \nonumber \\
\tau(r)&>&0  \; \; \mathrm{ for }\; \;  0\leq r \leq  \overline{\sigma} \equiv \sigma/2 \;
\end{eqnarray}
We denote by $y(r)$ the Yukawa potential created by a unit point charge. We have
$y(r)=\exp(-\alpha r)/r$ and for its Fourier transform $\widetilde{y}(k)=4 \pi/(\alpha^2 + k^2)$. 
The Yukawa potential $V_{\tau} \equiv \tau * y $ created by this distribution at point $\bf{R}$ is given by the convolution of $\tau$ and $y$ 
\begin{eqnarray}\label{V}
V_{\tau}(R)&=& \int_{0}^{\overline{\sigma}} dr \; 
             4 \pi r^2 \tau(r) I(R,r) \; ,\nonumber \\
I(R,r) &=&	\int \frac{d\Omega_{\bf{r}}}{4 \pi}  \; \frac{\exp(-\alpha \vert  \bf{R} -\bf{r})\vert)}{\vert \bf{R} -\bf{r}\vert} \; ,  
\end{eqnarray}
where $ d\Omega_{\bf{r}}$ denotes the infinitesimal spherical angle about vector $\bf{r}$.  The integral $I(R,r)$ reads \cite{Rosenfeld1,Rosenfeld2,Rosenfeld3,Rosenfeld4,Cai-Gilles}
\begin{eqnarray}\label{I}
I(R,r) &=& \frac{\sinh(\alpha r)}{\alpha r} \frac{\exp(-\alpha R)}{R} \; \; \mathrm{ for }\; \;  R>r \;, \nonumber \\
&=&\frac{\sinh(\alpha R)}{\alpha R} \frac{\exp(-\alpha r)}{r} \; \; \mathrm{ for }\; \;  r>R \;.
\end{eqnarray}
The potential $V_{\tau}(R)$ for $R<\overline{\sigma}$ is not
particularly useful but outside the sphere (i.e. for
$R>\overline{\sigma})$ one infers from equations\ (\ref{V}) and \
(\ref{I}) the remarkable result

\begin{eqnarray}
\label{V2}
V_{\tau}(R)&=& Q_{\tau} \frac{\exp(-\alpha R)}{R} \; , \\
Q_{\tau}&=& \int_{0}^{\overline{\sigma}} dr \; 
             4 \pi r^2 \tau(r) \frac{\sinh(\alpha r)}{\alpha r} \; .\label{Q}
\end{eqnarray}
 
As in the case of the Coulomb potential, the potential outside the spherical
distribution of charge $y(r)$ is still a Yukawa potential with the same
screening parameter $\alpha$ but with a different charge
$Q_{\tau}$. Since $\sinh(x)/x >0$ the effective charge $Q_{\tau}$ is
larger than the bare charge $\widetilde{\tau}(0)$ of the
distribution. Of course in the limit $\alpha \to 0$ one has $Q_{\tau}\to 1$ and
Gauss law is recovered.

We now prove the theorem 
\begin{equation}\label{inequaV}
- \frac{V_{\tau}(R)}{Q_{\tau}} + y(R) \geq 0 \; \; \forall R \;.
\end{equation}
For $R> \overline{\sigma}$ it is obvious; for $R < \overline{\sigma}$ we note that
from the expression\ (\ref{Q}) of  $Q_{\tau}$ and from equation\
(\ref{V}) it follows  that
\begin{equation}
V_{\tau}(R)-Q_{\tau} y(R)=
\int_{R}^{\overline{\sigma}} dr \; 4 \pi r^2 \tau(r) \frac{1}{\alpha r R} Z(r) \;,
\end{equation}
where $Z(r)=e^{-\alpha r}\sinh(\alpha R)-\sinh(\alpha r) e^{-\alpha R}$. Since
$Z(r)$ is a decreasing function of $r$ and $Z(R)=0$ we have $Z(r)\leq 0$ from which inequality\ (\ref{inequaV}) follows (note that $\tau(r)$ must be positive).
For a sufficiently regular distribution $\tau(r)$, the potential $V_{\tau}(R)$ is a bounded smooth function for $R <\overline{\sigma}$, in particular $V_{\tau}(R=0)$ is finite in general (see examples at the end of the appendix).

Clearly the Fourier transform $\widetilde{V}_{\tau}(k)=4 \pi \widetilde{\tau}(k)/(\alpha^2 + k^2)$ has not a definite sign in general although 
$\widetilde{V}_{\tau}(0)=4 \pi\widetilde{\tau}(0)\alpha^{-2} \; >0 $.

The interaction  $W_{\tau}(1,2)$ of two identical spherical
distributions of Yukawa charges centred at points $\bf{R}_{1}$ and
$\bf{R}_{2}$ is now given by  (with obvious notations)

\begin{eqnarray}\label{W}
W_{\tau}(1,2)&=& \int d1' \; d2' \; \tau(1,1') \; y(1',2') \; \tau(2,2') \nonumber \\
&=&\int d1'\; V_{\tau}(1,1') \; \tau(1',2) \; .
\end{eqnarray}
Also in this case Gauss theorem generalises easily. Indeed, for the case where the two distributions do not overlap, i.e. $R =
\vert  \bf{R}_{2} - \bf{R}_{1}\vert  >\sigma$, $V_{\tau}(1,1')$ can be replaced by $Q_{\tau} y(1,1')$ in equation\ (\ref{W}) and thus 
\begin{equation}
W_{\tau}(R) = Q_{\tau}^2 \; y(R) \; \; \mathrm{ for} \;  R>\sigma \; .
\end{equation}
Note that the Fourier transform
$\widetilde{W}_{\tau}(k)=\widetilde{\tau}^2(k) \;\widetilde{y}(k) \geq
0$ is positive definite contrary to $\widetilde{V}_{\tau}(k)$.  

In addition, we  have the theorem
\begin{equation}
- \frac{W_{\tau}(R)}{Q_{\tau}^2} + y(R) \geq 0 \; \; \forall R \;.
\end{equation} 
The proof is trivial :
\begin{eqnarray}
W_{\tau}(1,2)&=& \int d1' \; \tau(1,1') V_{\tau}(1',2) \nonumber \\
&\leq& Q_{\tau}\int d1'\; \tau(1,1') \; y(1',2) \equiv Q_{\tau} V_{\tau}(1,2)
\nonumber \\
&\leq& Q_{\tau}^2 \; y(1,2) \; .
\end{eqnarray}
The function $W_{\tau}(r)$ is in general bounded in the core and in
particular $W_{\tau}(0)$ is finite. This charge smearing process
provides an easy way to regularise the Yukawa potential $y(r)$ in the
core which preserves the positivity of the interaction. 
 
Finally, we give explicit expressions for $Q_{\tau}$, $V_{\tau}$
for $r < \overline{\sigma}$, and the value $W_{\tau}(0)$ in simple cases
relevant for Sect.\ \ref{OMF} . Thus for a surface distribution 
\begin{equation}
\tau_{\sigma}(r) = \frac{1}{\pi \sigma^2} \delta(r- \overline{\sigma}) \; ,
\end{equation}
one finds
\begin{eqnarray}\label{surf}
Q_{\tau_{\sigma}} & = & \frac{\sinh(\alpha\overline{\sigma}) }{\overline{\sigma}} \nonumber \\
V_{\tau_{\sigma}}(r) &=& \frac{\exp(-\alpha \overline{\sigma})}{\overline{\sigma}} \frac{\sinh(\alpha r)}{\alpha r} \; \; \mathrm{ for}\;  r< \overline{\sigma} \nonumber \\
W_{\tau_{\sigma}}(0) &=& Q_{\tau_{\sigma}}\; \frac{\exp(-\alpha \overline{\sigma})}{\overline{\sigma}} \; .
\end{eqnarray}
For a volume distribution
\begin{equation}
\tau_{\rho}(r) = \frac{6}{\pi \sigma^3} \Theta( \overline{\sigma} - r ) \; ,
\end{equation}
($\Theta$ step function) one has
\begin{eqnarray}\label{vol}
Q_{\tau_{\rho}} & = & \frac{3}{(\alpha \overline{\sigma})^3 } \;
( \alpha \overline{\sigma} \cosh( \alpha \overline{\sigma}) -
\sinh( \alpha \overline{\sigma})) \nonumber \\
V_{\tau_{\rho}}(r) &=& \frac{3}{\alpha^2 \overline{\sigma}^3}
\; 
\left( 1 - \left( 1 + \alpha \overline{\sigma}\right)  \exp\left( - \alpha  \overline{\sigma} \right) \; \frac{\sinh \left( \alpha r\right) }{\alpha r}\right) \; \; \mathrm{ for}\;  r< \overline{\sigma} \nonumber \\
W_{\tau_{\rho}}(0)&=& \frac{2}{\alpha^2 \overline{\sigma}^3} \;
\left( 
1 - \left(1 + \alpha\overline{\sigma}  \right)
\exp\left( - \alpha\overline{\sigma}\right) Q_{\tau_{\rho}}  
\right) \; .
\end{eqnarray}

\end{document}